\documentclass[aps,prb,twocolumn,superscriptaddress,showpacs,floatfix]{revtex4}

\usepackage{hyperref}
\usepackage{graphicx}

\begin{document}

\title{Numerical Study of Spin-1/2 $XXZ$ Model on Square Lattice from Tensor Product States}

\author{Pochung Chen}
\email{pcchen@phys.nthu.edu.tw} %
\affiliation{Department of Physics, National Tsing Hua University,
Hsinchu 30013, Taiwan}

\author{Chen-Yen Lai} %
\affiliation{Department of Physics, National Tsing Hua University,
Hsinchu 30013, Taiwan}%
\affiliation{Department of Physics and Astronomy, University of
California, Riverside, California 92521, USA}

\author{Min-Fong Yang}
\affiliation{Department of Physics,
Tunghai University, Taichung 40704, Taiwan}

\date{\today}

\begin{abstract}
By means of the recently proposed algorithm based on the tensor
product states, the magnetization process of the spin-1/2
anti-ferromagnetic $XXZ$ model on a square lattice is
investigated. In the large spin-anisotropy limit, clear evidence
of a first-order spin-flip transition is observed as an external
magnetic field is increased. Our findings of the critical field
and the discrete jumps in various local order parameters are in
good agreement with the quantum Monte Carlo data in the
literature. Our results imply that this algorithm can be an
accurate and efficient numerical approach in studying first-order
quantum phase transitions in two dimensions.
\end{abstract}


\maketitle


Numerical simulations are usually required in the theoretical
investigation on strongly correlated systems, because analytical
solutions are not available in most cases. Consequently,
developing accurate and efficient numerical tools becomes one of
the central issues in the understanding of quantum many-body
systems. Recently, based on an efficient representation of
two-dimensional system's wave function through a tensor network, a
series of new simulation algorithms has been achieved. In
particular, the infinite projected entangled-pair states (iPEPS)
algorithm~\cite{iPEPS} has been proposed and applied to various
interesting systems with
success.~\cite{Zhou08,Orus09,Li09,Jordan09,Bauer0905.4880} In this
approach, the ground-state wave function is described by the
so-called tensor product state (TPS)~\cite{TPS-1,TPS-2} or the
projected entangled-pair state (PEPS).~\cite{PEPS-1,PEPS-2} Taking
into account possible translational symmetry in the ground state,
such a tensor network can be simply represented by copies of a
small number of tensors even for systems on infinite lattices.
After optimizing these tensors under specific prescriptions, a
number of physical properties can be calculated from the optimized
TPS/PEPS.

By handling tensor-product wave functions in different manners,
schemes distinct from iPEPS algorithm have also been put
forward.~\cite{Gu08,Jiang08} A virtue of these approaches is that
they can be implemented with ease. In Ref.~\onlinecite{Gu08}, the
optimized TPSs are determined via direct variational approach,
where the variational energies of systems of very large sizes are
efficiently evaluated by means of the tensor renormalization group
(TRG) method.~\cite{Levin07,TRG-2} The expectation values of
physical quantities are then calculated from the optimized TPS
again under the TRG method. This algorithm has been tested for
several two-dimensional (2D) quantum spin models,~\cite{Gu08} and
the results agree well with previous findings. Alternatively in
Ref.~\onlinecite{Jiang08}, 
the ground states of a TPS form are obtained by using the
power method through iterative projections.
This approach can be considered as a generalization
of the 1D infinite time-evolving block decimation
(iTEBD) method~\cite{iTEBD} to the two dimensional cases.
After getting the ground states, the TRG method~\cite{Levin07} is
employed to calculate the expectation values of physical
observables. It is shown that accurate results for the Heisenberg
model on a honeycomb lattice can be reached under this
approach.~\cite{Jiang08}

Due to the simplicity and efficiency of the iTEBD and the TRG
algorithms, the approach proposed in Ref.~\onlinecite{Jiang08}
can become one of the promising numerical methods in studying
quantum many-body systems once its general validity is established.
Recently, it is shown that TPS/PEPS ansatz is suited to study
the first-order phase transition.~\cite{Orus09}
However, because of the difference in optimizing ground states and
in evaluating expectation values, one may wonder if
the combined iTEBD and TRG algorithm can determine the
first-order phase transitions to the same accuracy as
the iPEPS algorithm does.

In order to provide further benchmark on the 
performance of the combined iTEBD and TRG algorithm,
in this work we investigate the magnetization
process of the spin-1/2 anti-ferromagnetic $XXZ$ model on a square lattice.
Here the large spin-anisotropy case is considered, where
the existence of first-order spin-flip transitions in the
magnetization process has been established by means of quantum
Monte Carlo (QMC) simulations.~\cite{Kohno97,Yunoki02,Batrouni} We
find that various local order parameters defined below change
discontinuously at a critical field, which clearly indicates the
appearance of a first-order transition. Moreover, satisfactory
results of the critical field and the discrete jumps in the local
order parameters are be obtained as compared to the previous QMC
findings.~\cite{Yunoki02} Our present investigation suggests that
this 
combined algorithm should also be an effective
numerical method in studying first-order quantum phase transitions
in two dimensions.


Before presenting our results, it is instructive to sketch the
combined iTEBD and TRG algorithm employed here. We know that the
ground state can in principle be determined through the imaginary
time evolution for a given initial state $|\Psi_0\rangle$:
 $|\Psi_{\rm GS}\rangle =
 \lim_{\tau\rightarrow\infty}\exp(-H\tau) |\Psi_0\rangle /
 \|\exp(-H\tau) |\Psi_0\rangle \|$.
If, just like the present case, the model Hamiltonian can be
written as a sum of terms $h_{\langle i,j\rangle}$ involving only
pairs of nearest-neighboring sites $i$ and $j$, the Suzuki-Trotter
formula~\cite{Trotter} can be exploited to decompose the imaginary
time evolution operation into a product of two-site evolution
operators: $U_{\langle i,j\rangle} = \exp(-h_{\langle i,j\rangle}
\delta\tau )$, where $\delta\tau \ll 1$. It is also known that any
wave function can always be approximated in a TPS form. A possible
construction of TPS for systems on a square lattice is to attach a
rank-five tensor $[\Gamma_i]^s_{lrud}$ to each site $i$ and a
diagonal singular value matrix (hence a vector) $[\lambda_{\langle
i,j\rangle}]_l$ to each bond of nearest-neighboring sites $i$ and
$j$. Here $s$ is the physical index with $s=1,2$ for the present
spin-1/2 case, and $l,r,u,d(=1 \cdots D)$ denote the virtual bond
indices in four directions. In general, better representation of a
given wave function can be achieved by increasing the bond
dimension $D$. Taking into account the possible translational
symmetry in the ground state under shifts by two lattice sites
both in the $x$ and $y$ directions, the tensor network can be
simply represented by copies of tensors within a $2\times 2$ unit
cell. That is, we are left with four independent $\Gamma_i$
tensors and eight independent $\lambda_{\langle i,j\rangle}$
matrices. The action of a two-site evolution operator $U_{\langle
i,j\rangle}$ on such a TPS can be absorbed by performing a
singular value decomposition, and thus leads to an update of the
$\Gamma_i$, $\Gamma_j$, and $\lambda_{\langle ij\rangle}$
tensors.~\cite{Jiang08} When eight nearest-neighboring bonds
within the $2\times 2$  unit cell are all updated, a complete
iteration is achieved. After sufficient time of such updating
iterations, the optimized ground state of the TPS form can be
generated. 

Since evaluation of the expectation values for a TPS under the
most straightforward method is exponentially difficult, for a
complete numerical algorithm, an efficient way to do these
calculations for large systems must be also constructed. Here the
TRG approach~\cite{Levin07,Gu08} is employed. For any operator
that can be decomposed into product of local operators,
$\hat{O}=\prod_i \hat{O}_i$, evaluating $\langle \Psi_{\rm
GS}|\hat{O}|\Psi_{\rm GS}\rangle$ for the TPS ground state
$|\Psi_{\rm GS}\rangle$ is equivalent to compute the contraction
of a corresponding tensor network of $\mathbf{T}$ tensors. Within
such a tensor network, the rank-four tensor $\mathbf{T}^i$ at site
$i$ is defined as
\begin{equation}
  [\mathbf{T}^i]_{\bar{l}\bar{r}\bar{u}\bar{d}}
  = \sum_{ss^\prime} \langle s^\prime| \hat{O}_i |s\rangle [A^i]^s_{lrud}
   \left( [A^i]^{s^\prime}_{l^\prime r^\prime u^\prime d^\prime} \right)^*
\end{equation}
with
\begin{eqnarray}
   [A^i]^s_{lrud} &\equiv&
  \sqrt{ [\lambda_{\langle i,i-\hat{x}\rangle}]_l
  [\lambda_{\langle i,i+\hat{x}\rangle}]_r
  [\lambda_{\langle i,i+\hat{y}\rangle}]_u
  [\lambda_{\langle i,i-\hat{y}\rangle}]_d }  \nonumber \\
  & & \times [\Gamma_i]^s_{lrud} \; ,
\end{eqnarray}
where $|s\rangle$ represents the spin state at site $i$.
$i\pm\hat{x}$ and $i\pm\hat{y}$ denote the nearest neighbors of
site $i$ in the $x$ and $y$ directions, respectively. ${\bar
l}=(l,l^\prime)$ is the double bond index, and $\bar{r},
\bar{u},\bar{d}$ are similarly defined.
The tensor network of $\mathbf{T}$ tensors then can be coarse-grained in an
iterative fashion.\cite{Gu08,Levin07}
Each complete renormalization group (RG) step
reduces the size of the network by a factor of 2. The accuracy of
such a RG process is controlled by a cutoff $D_{cut}$ on the
double bond indices of the coarse-grained tensor. Therefore, to
evaluate the contraction of the tensor network of size
$2^{n+1}\times 2^{n+1}$, we need only perform $n$ RG steps. To sum
up, the TPS provides an efficient way to approximate the 
2D wave functions. The agreement between the actual wave
function and the represented TPS wave function can be improved
simply by increasing the bond dimension $D$. Besides, the TRG
approach serves as an efficient tool to evaluate the expectation
values for a TPS ground state of very large systems, where the
accuracy can be systematically improved by increasing the cutoff
$D_{\rm cut}$. In the present work
we consider the bond dimension up to $D=5$ and keep
$D_{\rm cut}\ge D^2$ to ensure the accuracy of the TRG
calculation.

\begin{figure}[t]
\includegraphics[width=3in]{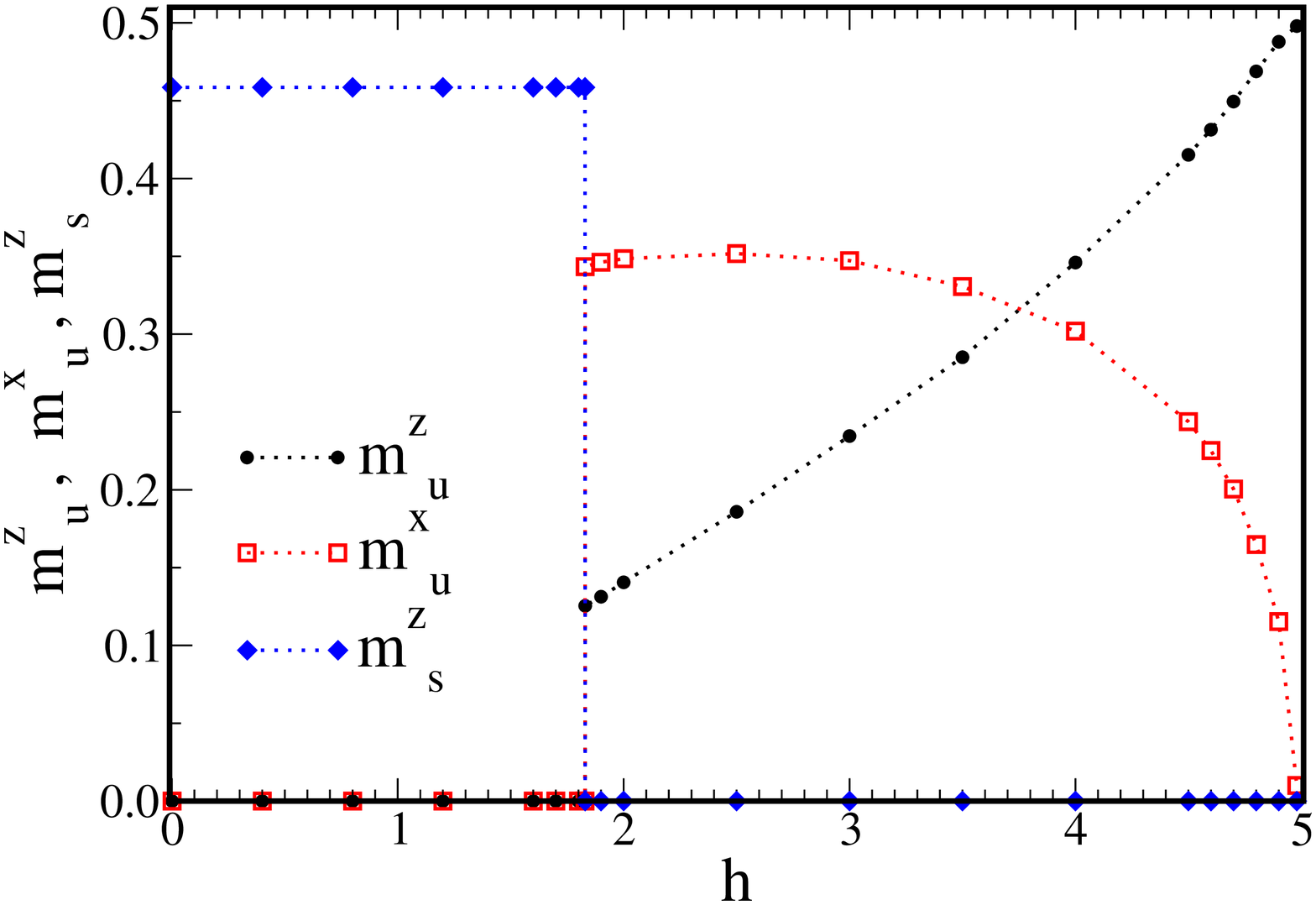}
\caption{(Color online)  Values of $m^z_u$, $m^x_u$, and $m^z_s$
for the ground state at $\Delta=1.5$ as functions of external
field $h$ for systems of size $2^7\times 2^7$ with $D=4$ and
$D_{\rm cut}=16$. } \label{fig:mag_Delta1.5}
\end{figure}

The general simulation procedure is described as follows. For a
given $h$ and $D$, we take a set of random $\Gamma$ and $\lambda$
tensors as our initial state $|\Psi_0\rangle$. While the initial
state may not have the spatial rotational symmetry, during the
imaginary time evolution, the evolved state will converge towards
a ground state which respects this expected symmetry. It hence
provides a self-consistent stability check for the algorithm. To
minimize the Trotter error, we usually start with
$\delta\tau=10^{-1}$ and gradually decrease it to
$\delta\tau=10^{-3}$ to ensure the convergence of the wave
function.

\begin{figure}[t]
\includegraphics[width=3in]{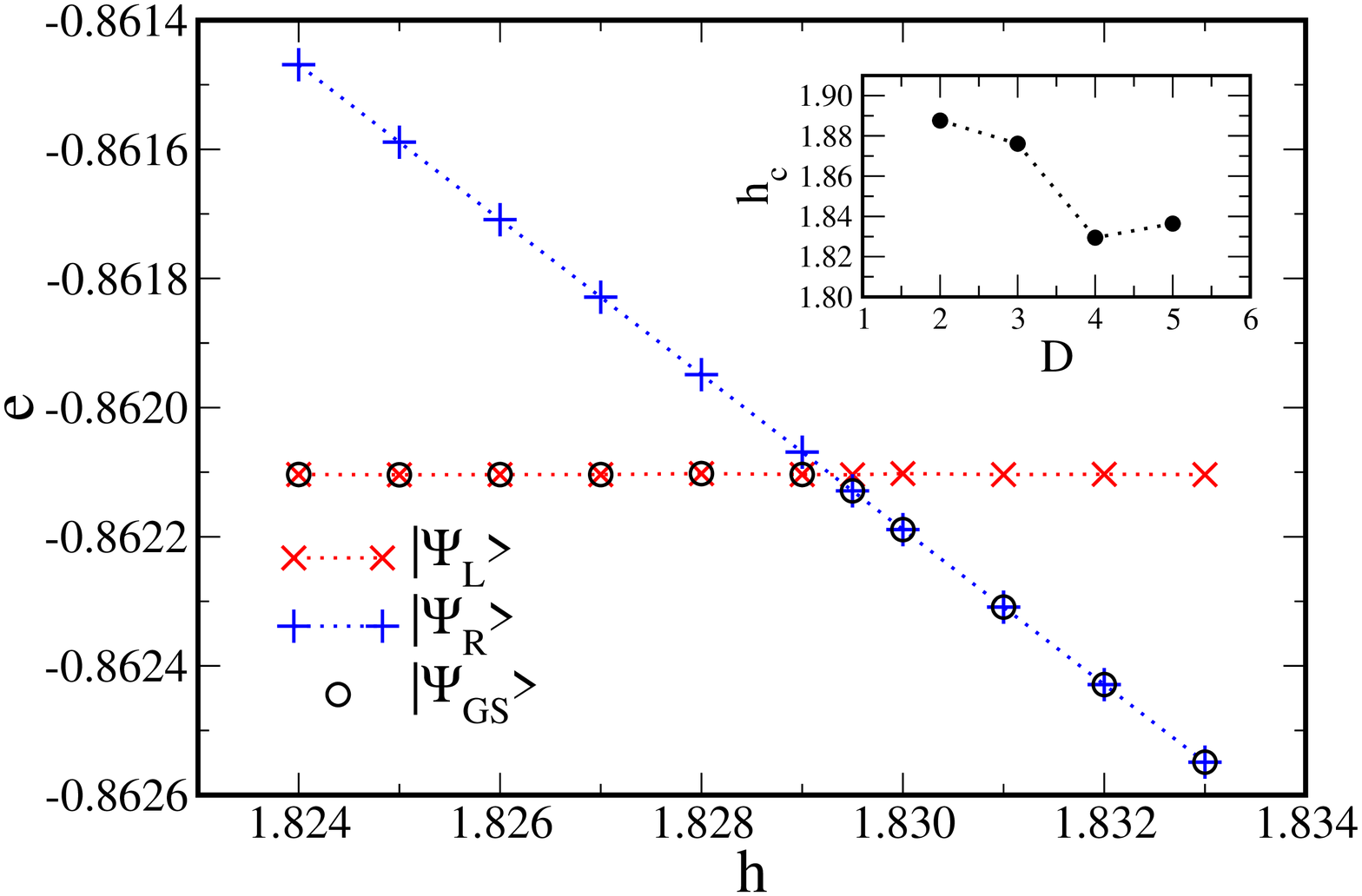}
\caption{(Color online)  Energies $e(h)$ per site for the
adiabatically evolved states $|\Psi_L(h)\rangle$ and
$|\Psi_R(h)\rangle$. The values for the ground state $|\Psi_{\rm
GS}(h)\rangle$ are denoted by open circles. Here $\Delta=1.5$ for
systems of size $2^7\times 2^7$ with $D=4$ and $D_{\rm cut}=16$.
The inset shows the critical field $h_c$ for various $D$ ($D_{\rm
cut}=16$ for $D\le 4$ and $D_{\rm cut}=25$ for $D=5$).
Dotted line is guide to eyes.} \label{fig:eng_Delta1.5}
\end{figure}


In the following, we present our numerical results for the
spin-1/2 $XXZ$ model with systems size $N=2^7\times 2^7$. In the
presence of an external magnetic field $h$ along $z$ direction,
the Hamiltonian of the $XXZ$ model is given by~\cite{note1}
\begin{equation}
H=J \sum_{\langle i,j \rangle} \left( -S^{x}_{i}S^{x}_{j} -
S^{y}_{i}S^{y}_{j} + \Delta S^{z}_{i}S^{z}_{j} \right) - h
\sum_{i} S^{z}_{i}, \label{eq:model}
\end{equation}
where $S^{\alpha}_{i}$ is the $\alpha$($=x,y,z$) component of the
spin-1/2 operator at site $i$, and $\langle i j \rangle$ runs over
all the nearest-neighboring pairs of spins at sites $i$ and $j$.
$J\equiv 1$ is the exchange coupling, and $\Delta$($\geq 0$) is an
anisotropic parameter. We focus our attention on the large
spin-anisotropy case of $\Delta=1.5$, where accurate QMC
calculations have been performed.~\cite{Yunoki02} The expectation
values of the $z$-component staggered magnetization $m^z_s \equiv
\sum_i \langle S^z_{i} \rangle e^{i{\bf Q}\cdot{\bf r_i}} /4$, the
uniform one $m^z_u \equiv \sum_i \langle S^z_{i} \rangle /4$, and
the $x$-component uniform magnetization $m^x_u \equiv \sum_i
\langle S^x_{i} \rangle /4$ for the ground states $|\Psi_{\rm
GS}(h)\rangle$ are shown in Fig.~\ref{fig:mag_Delta1.5}, where
${\bf Q}=(\pi,\pi)$ and the sum on $i$ runs over four sites within
the $2\times 2$ unit cell under consideration. Here we take the
bond dimension $D=4$ and the TRG cutoff $D_{\rm cut}=16$. We note
that results for $D=3,5$ are very similar to those for $D=4$ and
are thus not shown here. In the large spin-anisotropy limit with
$\Delta>1$, it is known that a first-order spin-flip transition
from a N\'{e}el-ordered phase to a spin-flopping phase will occur
as the external field $h$ increases from
zero.~\cite{Kohno97,Yunoki02,Batrouni} Crossing the critical field
$h_c$, the $z$-component staggered magnetization $m^z_s$ suddenly
drops to zero and the uniform part $m^z_u$ jumps to a nonzero
value. When $h>h_c$, $m^z_u$ increases monotonically and finally
reaches its saturated value ($m^z_u =1/2$) at
$h=h_s$[$=2(1+\Delta)$], while the staggered one $m^z_s$ remains
zero. A character of the spin-flopping states for $h_c<h<h_s$ is
the existence of finite $x$-component magnetization $m^x_u$, whose
value gives a measure of spin superfluidity (see below). When
spins become fully polarized in $z$ direction as $h$ approaches
$h_s$, $m^x_u$ will decrease to zero. As seen from
Fig.~\ref{fig:mag_Delta1.5}, our values of $m^z_s$, $m^z_u$, and
$m^x_u$ do show the expected results.

\begin{figure}[t]
\includegraphics[width=3in]{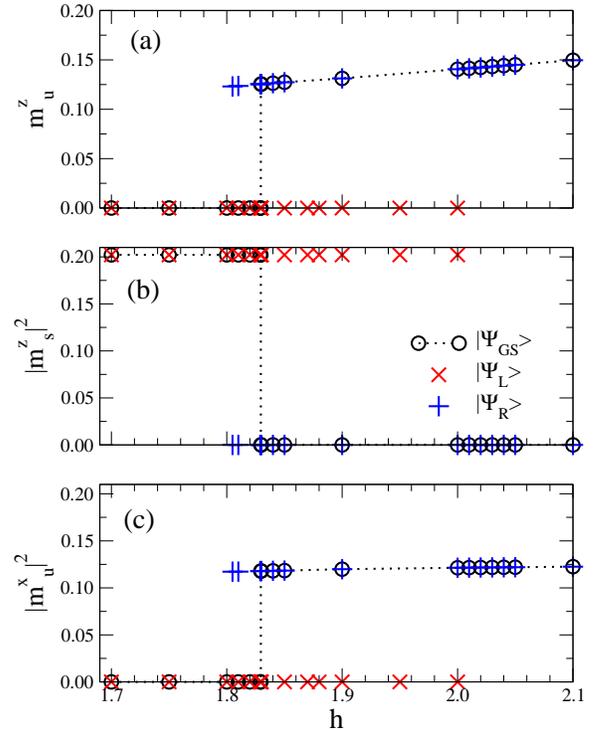}
\caption{(Color online)  Values of (a) $m^z_u$ (b) $|m^z_s|^2$ (c)
$|m^x_u|^2$ for the adiabatically evolved states
$|\Psi_L(h)\rangle$ and $|\Psi_R(h)\rangle$. The values for the
ground state $|\Psi_{\rm GS}(h)\rangle$ are denoted by open
circles. Here $\Delta=1.5$ for systems of size $2^7\times 2^7$
with $D=4$ and $D_{\rm cut}=16$.} \label{fig:hys_Delta1.5}
\end{figure}

Typically a first-order quantum phase transition comes from energy
level crossing in the ground state, and the crossing point gives
the critical value of the tuning parameter.~\cite{Sachdev:book} In
the present case, the relevant states should be the ground state
in the N\'{e}el-ordered phase with zero $m^z_u$ and that in the
spin-flopping phase with finite $m^z_u$. Here we simulate the
adiabatically evolved states $|\Psi_L(h)\rangle$ and
$|\Psi_R(h)\rangle$ starting from the computed ground states in
the N\'{e}el-ordered phase and in the spin-flopping phase,
respectively. That is, $|\Psi_L(h)\rangle$ ($|\Psi_R(h)\rangle$)
are determined starting from the ground state $|\Psi_{\rm
GS}(h_{\rm ini})\rangle$ for a given initial parameter $h_{\rm
ini}<h_c$ ($h_{\rm ini}>h_c$), and adiabatically increasing
(decreasing) $h$ in the Hamiltonian well beyond crossing the
critical field $h_c$. True ground states $|\Psi_{\rm
GS}(h)\rangle$ are the ones  with lower energies. The
corresponding energies $e(h)$ per site are shown in
Fig.~\ref{fig:eng_Delta1.5} for $D=4$ and $D_{\rm cut}=16$. We
find that the energies of $|\Psi_L(h)\rangle$ remain unchanged as
$h$ varies, while those of $|\Psi_R(h)\rangle$ are lowered as $h$
increases. This is expected since $|\Psi_L(h)\rangle$ should
behave like N\'{e}el-ordered states with zero $m^z_u$ and thus
their energies do not depend on the external field. However,
$|\Psi_R(h)\rangle$ should retain the spin-flopping character with
nonzero $m^z_u$, hence their energy expectation value $\langle H
\rangle$ for the Hamiltonian in Eq.~(\ref{eq:model}) should behave
as a decreasing function of $h$. Due to level crossing in these
two states, discontinuity in the first derivative of the ground
state energy appears. This again indicates the presence of a
first-order quantum phase transitions. From the crossing point in
Fig.~\ref{fig:eng_Delta1.5}, we find that $h_c\sim 1.829$, which
is quite close to the value estimated by QMC ($h_c\sim
1.83$).~\cite{Yunoki02} The dependence of $h_c$ on the bond
dimension $D$ is plotted in the insect of
Fig.~\ref{fig:eng_Delta1.5}. While the findings of $h_c$ for $D=2$
and 3 are somewhat higher than the QMC result, satisfactory values
can be obtained for larger $D$.

To show further evidence of a first-order transition between the
N\'{e}el-ordered phase and the spin-flopping phase, the results of
$m^z_u$, $|m^z_s|^2$, and $|m^x_u|^2$ for the adiabatically
evolved states $|\Psi_L(h)\rangle$ and $|\Psi_R(h)\rangle$, and
the corresponding values for the ground state $|\Psi_{\rm
GS}(h)\rangle$ are displayed in Fig.~\ref{fig:hys_Delta1.5}. We
find that the values of $m^z_u$, $|m^z_s|^2$, and $|m^x_u|^2$ for
the ground state are all discontinuous at $h_c$. Moreover, all the
results for the adiabatically evolved states $|\Psi_L(h)\rangle$
and $|\Psi_R(h)\rangle$ show clearly the hysteresis behaviors.
These facts strongly support the presence of a first-order
transitions. The discrete jumps at $h_c$ for $m^z_u$ and
$|m_s^z|^2$ are $m^z_{u,c}\sim 0.125$ and $|m^z_{s,c}|^2 \sim
0.202$,
%
%
respectively. Both of them agree with the QMC results reported in
Ref.~\onlinecite{Yunoki02}: $m^z_{u,c}\sim 0.11$ and
$|m^z_{s,c}|^2 \sim 0.20$.~\cite{note2} As mentioned before, in
the spin-flopping phase for $h_c<h<h_s$, there exists spin
superfluidity which can be characterized by nonzero spin stiffness
$\rho_s$ (or the superfluid density in the corresponding hard-core
boson model). It is shown that $\rho_s$ changes discontinuously at
the first-order spin-flip transition.~\cite{Yunoki02} To the best
of our knowledge there is no straightforward way to calculate the
spin stiffness $\rho_s$ within the TPS framework. Instead, the
square of the $x$-component magnetization $m^x_u$ are evaluated,
which can be related to the density of Bose condensate in the
corresponding hard-core boson model.~\cite{note3} Thus the fact of
non-vanishing $|m^x_u|^2$ also implies the existence of spin
superfluidity. As seen from Fig.~\ref{fig:hys_Delta1.5}(c),
similar to the behavior of $\rho_s$ observed in the QMC study,
$|m^x_u|^2$ has also a discrete jump at $h_c$, which is of
magnitude $|m^x_{u,c}|^2 = 0.118$.
%


In summary, the first-order spin-flip transition
of the spin-1/2 $XXZ$ model with large spin
anisotropy can be detected under the combined iTEBD and TRG
algorithm proposed in Ref.~\onlinecite{Jiang08}. Good agreement
with the accurate QMC calculations can be
obtained by using merely moderate bond dimension $D$ and the TRG
cutoff $D_{\rm cut}$.
This demonstrates that the current formalism
will be a competitive numerical method
to determine particularly first-order quantum phase transition
in two dimensions, with the simplicity and efficiency as its advantage.
We note, however, that further
investigations are necessary to establish its general validity,
and to explore its relative performance as compared to other
TPS/PEPS-based approaches.

C.-Y. Lai and P. Chen thank the support from the National Science
Council of Taiwan under Contract No. NSC 95-2112-M-007-029-MY3.
M.-F.Yang acknowledges the support by the National Science Council
of Taiwan under Grant No. NSC 96-2112-M-029-004-MY3. This work is
supported by NCTS of Taiwan.


\begin{thebibliography}{99}
\bibitem{note1}
The Hamiltonain described in Eq.~(\ref{eq:model}) is equivalent to
that of the conventional antiferromagnetic $XXZ$ model by the
following unitary transformation: rotating the spins along the $z$
axis with an angle $\pi$ (thus $S^x_i \rightarrow -S^x_i$ and
$S^y_i \rightarrow -S^y_i$) on the sublattice consisting of, say,
even sites.
\bibitem{note2}
Within the N\'{e}el-ordered phase, one can show that $S({\bf Q})/N
=|m_s^z|^2$ in the thermodynamic limit, where $S({\bf Q}) \equiv
(1/N)\sum_{i,j} \langle {\bf S}_{i} \cdot {\bf S}_{j} \rangle
e^{i{\bf Q}\cdot({\bf r_i}-{\bf r_j})}$ is the spin structure
factor. The QMC data for $S({\bf Q})/N$ in the thermodynamic limit
can be read off from Fig.~2(b) of Ref.~\onlinecite{Yunoki02}.
\bibitem{note3}
For the spin-flopping states with the in-plane magnetization along
the $x$ direction, by using the relations between boson operators
and the spin ones (see Ref.~\onlinecite{Batrouni}), we have
$\rho_0 =|m_u^x|^2$ in the thermodynamic limit. Here $\rho_0$ is
the Bose condensate fraction in the hard-core boson model given by
$\rho_0 = (1/N^2)\sum_{i,j} \langle a^\dag_{i} a_{j} \rangle$.
\end{thebibliography}
\end{document}